# AI-Assisted Workflow Optimization and Automation in the Compliance Technology Field

Zhen Zhong
Senior Data Engineer, Graduate School of Arts & Sciences, Georgetown University, Washington, DC 20001, United States

***Abstract*—Against the backdrop of digital transformation and stricter regulation, enterprise compliance work demands higher efficiency and accuracy. The auxiliary compliance process has become an important entry point for optimizing the compliance system due to its strong transactional nature and high degree of repetition. This study focuses on the process characteristics of auxiliary compliance work, sorts out its structural composition and organizational mechanism, proposes an optimization path with process reengineering, system modeling, and technology integration as the core, and focuses on exploring the collaborative application of key technologies such as RPA, rule engine, and semantic recognition in process automation. Research suggests that the systematic optimization and intelligent upgrading of auxiliary processes will help build a modern compliance operation system that is responsive, efficient, structurally clear, and risk controllable.**

**Keywords—*Compliance technology; auxiliary process; process optimization; automation***

## I. Introduction

With the continuous escalation of compliance requirements worldwide, multinational corporations face growing complexity in aligning operations with diverse regional regulations. Existing studies have applied robotic process automation (RPA), rule engines, and semantic recognition to improve efficiency in transactional compliance, particularly within finance and healthcare sectors in Europe and North America. However, these approaches remain limited by dependence on structured data, poor adaptability to rapidly changing regulations, and weak cross-industry transferability.

Most research focuses on static rule-based automation, while practical compliance increasingly demands dynamic adjustment to new laws and contextual anomalies. Organizations, therefore, struggle to maintain both precision and agility. Future advancements should integrate AI-driven anomaly detection, real-time regulatory monitoring, and adaptive policy learning. Expanding these capabilities beyond finance and healthcare to manufacturing, logistics, and telecommunications could enhance global compliance resilience. Auxiliary compliance processes, with their high frequency and standardization, offer a strategic foundation for achieving this integration. This study introduces an integrated compliance automation framework that combines adaptive threshold control, RPA orchestration, and semantic rule modeling—an approach not yet systematically implemented in prior literature.

## II. Process Characteristics of Auxiliary Compliance Work

### A. Logical Division of Task Types and Execution Methods

In international corporate compliance management, auxiliary compliance tasks are generally categorized into data collection and entry, document verification and archiving, standardized report generation, rule matching, and anomaly flagging. These tasks typically rely on clearly defined rules and fixed-step operational patterns. For example, in anti-money laundering (AML) compliance within the financial sector, the backend system may pull data from multiple sources to match customer identity information, followed by compliance list screening carried out by either human operators or automated tools. In the medical data processing domain, compliance teams perform formatting and privacy redaction of electronic medical records in accordance with regulations such as HIPAA. These tasks exhibit a high degree of divisibility and repeatability, making them well-suited for partial or complete automation to reduce human error and accelerate processing [1].

### B. Dimensions for Identifying Typical Processes and Operational Bottlenecks

The typicality of auxiliary compliance processes can be identified through dimensions such as task frequency, data structure complexity, cross-departmental interaction volume, and the rate of regulatory change. In international cross-border payment compliance reviews, high-frequency transaction screening and cross-system data matching often become bottlenecks due to inconsistent data formats and lengthy information transmission chains. In the energy sector's carbon emissions compliance audits, bottlenecks are concentrated in the parsing and standardization of unstructured report texts. By identifying these dimensions, it becomes possible to pinpoint the process segments most amenable to technological intervention for efficiency improvement, thereby providing targeted guidance for subsequent process reengineering and automation deployment.

### C. Process Organization Characteristics of Manual Processing and Data Structure

Under traditional models, auxiliary compliance processes often rely on human operators for task scheduling and result confirmation, with data handling structures typically characterized by decentralization and non-real-time processing. For example, in international insurance claim compliance reviews, staff must switch between multiple business systems to obtain complete information, resulting in extended task completion times and difficulties in ensuring data consistency.

In the telecommunications sector's user privacy compliance management, manual workflows frequently involve handling semi-structured data, such as customer communication logs and transaction receipts, which increases data cleansing costs and limits real-time risk response capabilities. These organizational characteristics underscore the necessity of strengthening data structure standardization and system interoperability in process design, laying the groundwork for subsequent automated execution.

## III. Systematic Design for Assisting Workflow Optimization

### A. Underlying Logic of Task Decomposition and Process Reengineering

In cross-border KYC/AML scenarios, auxiliary compliance workflows can be abstracted as a directed acyclic graph G=(V,E)， where each task $v \in V$ is decomposed into atomic operations $v \to \{a_k\}$ and designated for either human (H) or robotic (R) execution, with $x_v \in \{H,R\}$. The optimization objective is defined as [Formula (1)]:

$$\min J = \alpha \cdot T(G,x) + \beta \cdot \varepsilon(G,x) + \gamma \cdot C(G,x) \quad (1)$$

subject to [Formula (2)]:

$$s(u) + t_u \leq s(v)(u,v) \in E、\ \tau(v) \leq \hat{\tau}(SLA) \quad (2)$$

and compliance predicate [Formula (3)]:

$$\Phi(d) = true(GDPR / HIPAA) \quad (3)$$

Incoming data are standardized through a mapping function (d)=σ(φ(d)) before entering the pipeline, ensuring that privacy processing $\pi(d) = mask(d,k)$ and kkk-anonymity ($k \geq k0$). requirements are met [2].

A rule engine $R=\{r_i\}$ drives decision-making, where:

$$r_i : p_i(d) \geq \theta \Rightarrow action \ldots \quad (4)$$

and the probability $p_i$ is computed by a semantic model. The scheduler allocates tasks between RPA and human reviewers based on critical path analysis and queue load: when $p_i<\theta$ or the incremental risk $\Delta risk(v)>\delta$, $x_v$ is switched to H and a secondary audit is triggered. In the objective function, T(G,x) represents the total task completion time determined by workflow structure G and execution mode x; ε(G,x) denotes the cumulative error rate from both automated and manual tasks; C(G,x) is the overall operational cost, including human labor and system resource usage. Coefficients α,β,γ define their relative weighting according to compliance priorities—speed, accuracy, or cost efficiency. The constraint s(u)+t u≤s(v) ensures task sequencing consistency, preventing logical overlap in dependent operations. τ(v)≤τ^enforces service-level compliance for each task, while the predicate Φ(d)=true guarantees that all data handling steps conform to privacy laws such as GDPR and HIPAA. The parameter θ is a dynamic confidence threshold determining whether decisions are automated or escalated, and Δrisk(v) quantifies deviations in risk estimates used to trigger human intervention. This mathematical formalization provides a unified framework for optimizing efficiency, accuracy, and compliance integrity across heterogeneous processes. For example, in a European retail bank's account opening process, the sequence of passport OCR → entity parsing → sanctions list screening → risk stratification can be reengineered under this model, resulting in significant reductions in both processing time T and error rate ε, while meeting SLA requirements. Unlike prior work that isolates automation from compliance logic, this framework unifies workflow optimization, real-time feedback control, and rule-based governance into a single adaptive architecture.

### B. Process Construction Path Based on Scenario Logic and Modeling Rules

Following the completion of task decomposition and process reengineering logic design, the construction of scenario logic and modeling rules becomes the key step in transforming the abstract model into an executable workflow. This process involves mapping the atomized task units and execution nodes to concrete business scenarios, ensuring that the workflow achieves automation and efficiency while meeting regulatory requirements [3]. For example, in the compliance review process for account opening at a UK retail bank, the event of a customer submitting an identity document triggers the system to call the OCR module for image parsing. The parsed result is immediately sent to the sanctions list screening engine, which applies predefined rules to determine the presence of high-risk matches. When the match confidence falls below 0.90, the workflow automatically switches to the manual review path and simultaneously generates a complete processing record in the backend to ensure subsequent compliance with regulatory traceability requirements.

The formulation of modeling rules must balance the mandatory nature of legal provisions with the operational flexibility of the institution, typically comprising three categories: legal provisions, internal control rules, and model thresholds, all unified into executable policies. For instance, when processing electronic medical records, a large U.S. healthcare group embeds the HIPAA-mandated PHI field list into the rule engine as the legal layer rules, ensuring that all identified sensitive fields are automatically redacted before data transmission. At the internal control layer, fields deemed highly sensitive must undergo verification through a dual-review mechanism; at the model threshold layer, the system requires the named entity recognition model to achieve a confidence score of no less than 0.92 before automatically masking the identified entity. All rules must pass static conflict detection and priority sequencing prior to deployment to prevent execution conflicts. Through unified interface calls, the workflow can operate stably across systems, with historical case playback and canary testing used to validate the feasibility and accuracy of the strategy. This approach establishes a compliance workflow system that is scalable, interpretable, and capable of rapid iteration.

### C. Process Collaboration Mechanism and Feedback Loop Design

The collaboration and feedback loop are driven by observable metrics to enable adaptive adjustment of thresholds and resources. The core approach is to continuously monitor key runtime indicators—end-to-end latency T, false positive rate (FPR), false negative rate (FNR), and backlog volume B—and to consolidate them into a single weighted-deviation objective for parameter tuning [Formula (5)]:

$$(J = \sum w_k (m_k - m_k^{\backslash *})^2) \tag{5}$$

Node load can be controlled with a single constraint:

$$\rho = \lambda / (c\mu) \le \rho_{\max} \tag{6}$$

where, $\lambda$ is the arrival rate, $\mu$ is the service rate, and c is the number of concurrent channels (including both RPA bots and human reviewers). Routing uses a risk score threshold $\theta$: when $s \le \theta$, processing is automated; when $s > \theta$, the case is routed to manual review. Both $\theta$ and the concurrency c of each node are incrementally adjusted by the controller following the principle of "the greater the deviation, the stronger the adjustment," while remaining within feasibility and compliance constraints such as SLA adherence, audit logging, and data minimization. For example, in the KYC account opening process at a UK retail bank, the sanctions list screening node recorded a load of $\rho=0.86$ during peak hours, exceeding the alert threshold of 0.80 [4]. The system first increased concurrency for that node from 8 to 12 and temporarily reduced OCR node concurrency to maintain the overall capacity cap. Within the following 10 minutes, the FPR rose to 3.1 per cent, prompting the controller to fine-tune $\theta$ from 0.88 to 0.84, increasing the manual review proportion to 22 per cent, bringing the false positive rate back within the 2.0 per cent target range and clearing the backlog. Data distribution drift is monitored using the Population Stability Index (PSI). When the upper PSI threshold is triggered, the system switches to "conservative mode", freezing $\theta$ adaptation, allowing only capacity-side fine-tuning, and increasing the sampling rate for high-confidence cases. All routing and parameter changes are simultaneously recorded in the audit log (sample ID, rule version, evidence chain) to ensure full traceability and reproducibility.

## IV. Key Technology Path for Compliance Process Automation

### A. Task Substitution and Process Execution Based on RPA

Building on the above foundations of task atomization and scenario–rule executable design, RPA functions as the "executor", transforming abstract nodes into controllable automated actions and integrating feedback loop parameters (thresholds and concurrency) to form an adaptive operating layer. The implementation path uses BPMN/DMN as the primary framework, registering task modules such as "document parsing, list screening, risk categorization, and audit archiving" into the orchestrator, with clearly defined input/output contracts and error semantics. Each task module is linked to a pool of unattended bots and manual review seats, with the scheduler dynamically allocating concurrency c according to the node load Formula (6), and using the routing threshold $\theta$ to control the proportion of automated versus manual processing (risk score $s \le \theta$ proceeds automatically, while scores above the threshold are routed to manual review).

Before data enters the RPA, minimization and masking are applied. Idempotent keys (sample_id + hash) ensure "exactly-once" execution, with failed executions applying exponential backoff and entering a dead-letter queue for manual takeover. Cross-system calls use short-lived tokens and a secure key vault to limit permission scope and access duration. To support iterative updates of rules and models, all bots are packaged and versioned as container images, with sandbox replays of historical cases comparing three key indicators—processing latency, false positive/false negative rates, and backlog volume—before gradual rollout through gray traffic. During runtime, telemetry continuously reports node latency, queue lengths, and error codes, enabling the controller to fine-tune $\theta$ and c, while writing parameters and evidence chains into an immutable audit log. For example, in the KYC account opening process of a UK retail bank, the "passport upload" event triggers an unattended RPA to call an OCR service, outputting document layout structures and entity candidates. These entities are matched against sanctions lists in the rule engine; if the list similarity score is 0.86 and below the threshold $\theta=0.90$, the case follows the automated approval branch and generates an audit record. During peak hours, when the sanctions list screening node load $\rho=0.83$ approaches its limit, the orchestrator increases concurrency for that node from 8 to 12 and temporarily reduces OCR node concurrency to maintain the overall capacity cap, bringing end-to-end latency back within 180s in 10 minutes. If the false positive rate briefly rises above 3 per cent, the controller fine-tunes $\theta$ to 0.84, increasing the manual review share until the FPR returns to target range, after which the parameters automatically revert.

In the exception path, anti-crawling and document tamper checks are performed by a security-aware bot before processing, triggering high-risk flags and secondary review if needed. Cross-border address verification is handled via a geosanction API call by the bot, which retrieves real-time lists and caches signed snapshots to meet traceability requirements. To enhance explainability, each automated approval or rejection outputs a "data version – rule version – model threshold – evidence snippet" tuple for audit sampling and regulatory inquiries. Fig. 1, UK Retail Bank KYC Automation Orchestration Diagram (Event → Task Module → Routing → Feedback Loop Adjustment) may be cited to illustrate the interactions between the orchestrator, bot pool, manual stations, and monitoring/logging components.

### B. Mechanism for Structured Conversion of Text Data

For unstructured text in cross-border compliance contexts (such as scanned documents, PDFs, emails, and chat logs), the conversion mechanism is implemented as a traceable pipeline consisting of "ingestion – parsing – extraction – alignment – validation – storage", integrated with the previously described threshold and concurrency control. In the ingestion stage, fingerprint deduplication and language detection are performed, generating idempotent keys and page coordinate indexes, with sensitive segments undergoing minimization and partial masking before processing. In the parsing stage, layout analysis is combined with adaptive OCR to differentiate between paragraphs, tables, and key-value areas, while a document-type classifier routes passports, bank statements, medical records, and other documents to the corresponding template families or template-free extractors.

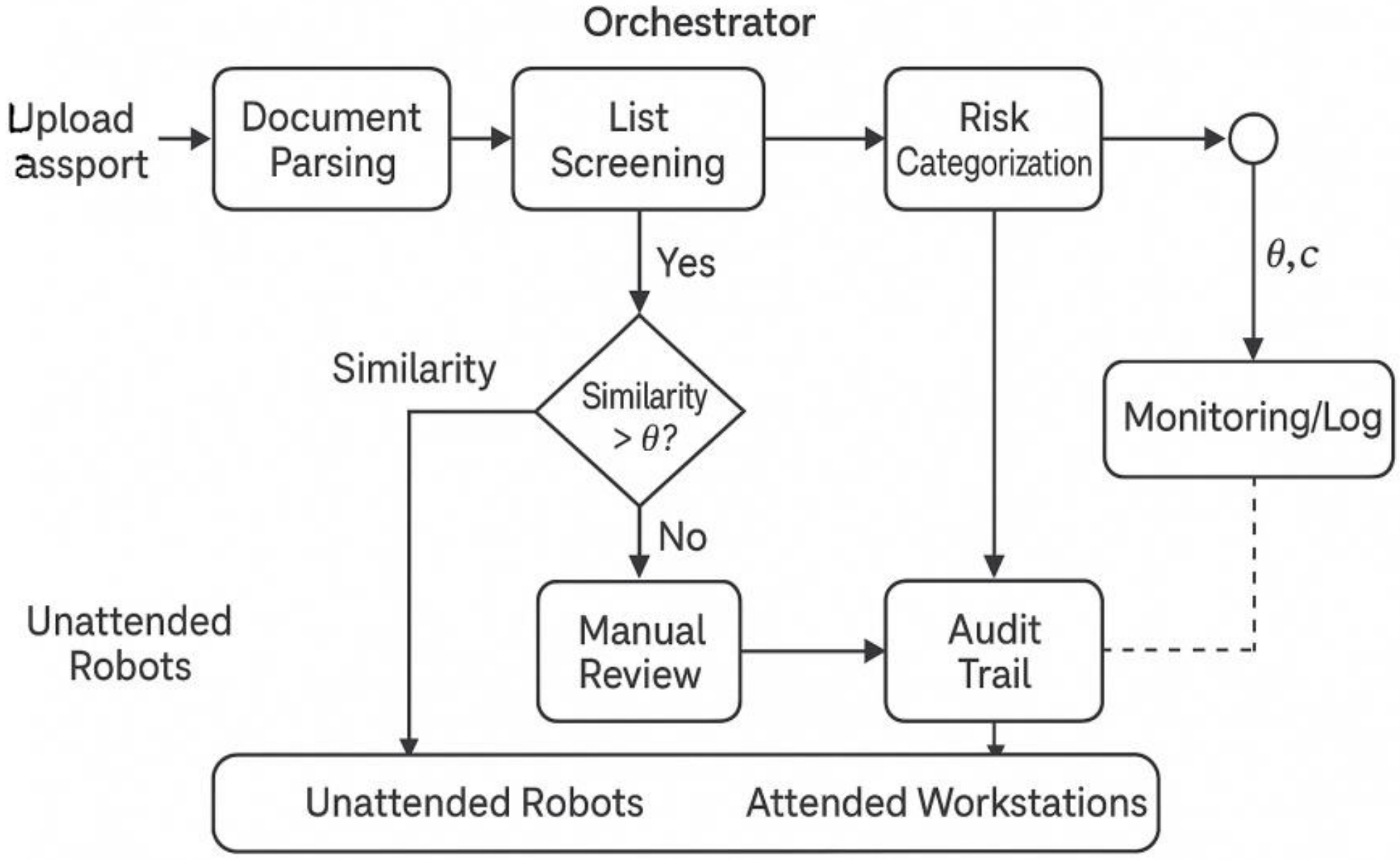

Fig. 1. UK retail bank KYC automation orchestration diagram.

The extraction stage operates under a hybrid "rule + model" paradigm: regular expressions and dictionaries ensure high-precision identification of strongly formatted fields such as ID numbers, IBAN, SWIFT codes, and CPT/ICD codes, while named entity recognition and relationship extraction models cover names, addresses, organizations, dates, and transactional semantics. These outputs include field-level confidence scores and evidence fragments (page number, bounding box, matching path). In the alignment stage, candidate entities are mapped to a unified business schema and external standards (ISO-8601 dates, ISO-3166 country/region codes, ISO-4217 currency codes, LEI/SSN placeholders), with cross-document merging achieved through entity resolution. Blocking keys and similarity metrics within the threshold range trigger manual review [5].

During validation, both field-level and sample-level thresholds are applied for consistency checks and compliance predicate evaluations (data minimization, retention periods, cross-border transfer flows). Exceptions are routed to compensating transactions and dead-letter queues while generating an auditable evidence chain. In the storage stage, versioned JSON Schemas and event sourcing are used to record evidence packages, rule versions, and model versions, enabling playback and incremental deployment.

For example, in the de-identification process of electronic health records (EHR) at a U.S. hospital group, the system ingests mixed HL7/PDF records, uses layout analysis to locate tables and free-text areas, and employs PHI detection models in combination with dictionary rules to output candidates for names, addresses, phone numbers, and geographic identifiers along with confidence scores. When an address confidence score falls below the set threshold, the case is routed to a manual workstation for verification, while high-confidence fields are directly masked or replaced with pseudo-random values. All changes are logged in an immutable audit record containing the sample ID, field, evidence fragment, and audit signature. Field-level false positive and false negative statistics are fed back to the monitoring dashboard to adjust sampling rates and thresholds online, ensuring high-quality, interpretable structured output under GDPR/HIPAA constraints [6].

### C. Integrated Design of Rule Engine and Automated Decision-Making Mechanism

The integration approach uses a three-layer rule stack—"legal hard constraints – internal policies – model scoring"—to drive decisions. DMN/decision tables and executable DSL are employed to define conditions, actions, and priorities. At runtime, external lists, thresholds, and black/white lists are mounted as snapshot versions to ensure replayability and traceability. The decision process consists of pre-check, rule matching, conflict resolution, and action orchestration: the pre-check layer verifies data completeness and timeliness; the matching layer triggers based on a combination of static rules and dynamic thresholds; the conflict resolution layer applies a priority order of "regulatory rejection > risk block > business approval"; and the action layer works with the orchestrator to issue approvals, rejections, or "enhanced verification" directives [7].

The risk score s output from the model and the current operating threshold θt act as soft evidence in the rules, while any "hard" regulatory matches override the model's decision. Change management uses policy versioning and case replay, with all new rules required to pass unit test suites, historical sample shadow runs, and gray (canary) releases. Online monitoring of FPR/FNR, average decision latency, and trigger rates ensures stability.

For example, in a PSD2/SCA and AML scenario at a European payment institution, the rule table specifies that "originating from a high-risk country + abnormal device fingerprint + transaction amount exceeding the SEPA threshold" triggers "enhanced identity verification". If a sanctions list hit occurs simultaneously, the transaction is rejected outright. When the daily false positive rate increases, operations adjust only θt and the sanctions list similarity threshold without altering the legal layer rules. The decision explanation is returned to the audit system in the format "active rule ID – evidence fragment – version number", ensuring interpretability, controllability, and iterative capability [8].

## V. Conclusion

Against the backdrop of increasingly stringent global compliance requirements, building an automation system centered on process reengineering, scenario logic modeling, and key technology integration can simultaneously improve execution efficiency and strengthen risk control capabilities. RPA-based task substitution, structured conversion of text data, and the rule engine's automated decision-making mechanism—supported by feedback loops and collaborative scheduling—enable balanced node load, adaptive thresholds, and full traceability. Validated through cases in UK retail banking and healthcare institutions in Europe and North America, the system has demonstrated strong stability and scalability in controlling processing latency, false positives, and false negatives, offering a replicable technical pathway for cross-industry and cross-regional compliance operations.